\pdfoutput=1

\documentclass[11pt]{article}
\usepackage{booktabs}
\usepackage{tabularx}
\usepackage{amsmath}
\usepackage{tcolorbox}
\usepackage[table]{xcolor}    
\usepackage{booktabs}         
\usepackage{etoolbox}
\usepackage{colortbl}
\usepackage{xcolor}
\usepackage{array}
\usepackage[flushleft]{threeparttable} 
\renewcommand{\arraystretch}{1.1}      
\newcolumntype{L}{>{\raggedright\arraybackslash}X}  
\newcolumntype{C}{>{\centering\arraybackslash}X}
\newcommand{\ccrrow}{\rowcolor{red!10}}      
\newcommand{\ccbrow}{\rowcolor{blue!10}}     
\newcommand{\ccgrow}{\rowcolor{green!10}}    
\newcommand{\ccyrow}{\rowcolor{purple!10}}   
\newcommand{\ccmrow}{\rowcolor{yellow!10}}   

\newcommand{\ccr}[1]{\textcolor{red}{#1}}      
\newcommand{\ccb}[1]{\textcolor{blue}{#1}}     
\newcommand{\ccg}[1]{\textcolor{green!60!black}{#1}}    
\newcommand{\ccm}[1]{\textcolor{purple}{#1}}   
\newcommand{\ccy}[1]{\textcolor{yellow!80!black}{#1}}   

\robustify{\ccr}
\robustify{\ccb}
\robustify{\ccg}
\robustify{\ccy}
\robustify{\ccm}

\newcolumntype{\ccrcol}{>{\columncolor{red!10}}c}
\newcolumntype{\ccbcol}{>{\columncolor{blue!10}}c}
\newcolumntype{\ccgcol}{>{\columncolor{green!10}}c}
\newcolumntype{\ccycol}{>{\columncolor{yellow!10}}c}
\newcolumntype{\ccmcol}{>{\columncolor{magenta!10}}c}

\usepackage{acl}

\usepackage{times}
\usepackage{latexsym}
\newcommand{\name}{\textbf{\textit{M3Retrieve}}}

\usepackage[T1]{fontenc}

\usepackage[utf8]{inputenc}
\usepackage[table]{xcolor}
\usepackage{booktabs,tabularx}
\newcolumntype{L}{>{\raggedright\arraybackslash}X}

\usepackage{microtype}

\usepackage{inconsolata}

\usepackage{graphicx}

\title{\name: Benchmarking Multimodal Retrieval for Medicine}

\author{
    \textbf{Arkadeep Acharya}$^{1}$\thanks{Equal contribution. Contact author: \href{mailto:acharyarka17@gmail.com}{acharyarka17@gmail.com}, \href{mailto:akash_2321cs19@iitp.ac.in}{akash\_2321cs19@iitp.ac.in}} 
    \thanks{Currently working at  IBM Research, Bangalore, India.}
    \quad
    \textbf{Akash Ghosh}$^{1}$\footnotemark[1] \quad
    \textbf{Pradeepika Verma}$^{1}$ \\
    \textbf{Kitsuchart Pasupa}$^{2}$ \quad
    \textbf{Sriparna Saha}$^{1}$ \quad
    \textbf{Priti Singh}$^{1}$ \\
    \\
    $^{1}$Indian Institute of Technology Patna, India \\
    $^{2}$King Mongkut's Institute of Technology Ladkrabang, Thailand
}

\begin{document}
\maketitle
\begin{abstract}
With the increasing use of Retrieval-Augmented Generation (RAG), strong retrieval models have become more important than ever. In healthcare, multimodal retrieval models that combine information from both text and images offer major advantages for many downstream tasks such as question answering, cross-modal retrieval, and multimodal summarization, since medical data often includes both formats. However, there is currently no standard benchmark to evaluate how well these models perform in medical settings. To address this gap, we introduce \name, a Multimodal Medical Retrieval Benchmark. \name , 
 spans 5 domains,16 medical fields, and 4 distinct tasks, with over 1.2 Million text documents and 164K multimodal queries, all collected under approved licenses. We evaluate leading multimodal retrieval models on this benchmark to explore the challenges specific to different medical specialities and to understand their impact on retrieval performance. By releasing \name, we aim to enable systematic evaluation, foster model innovation, and accelerate research toward building more capable and reliable multimodal retrieval systems for medical applications. The dataset and the baselines code are available in this github page \url{https://github.com/AkashGhosh/M3Retrieve}.

\end{abstract}
\section{Introduction}
\begin{figure*}[htbp]
    \centering
    \includegraphics[width=0.9\linewidth]{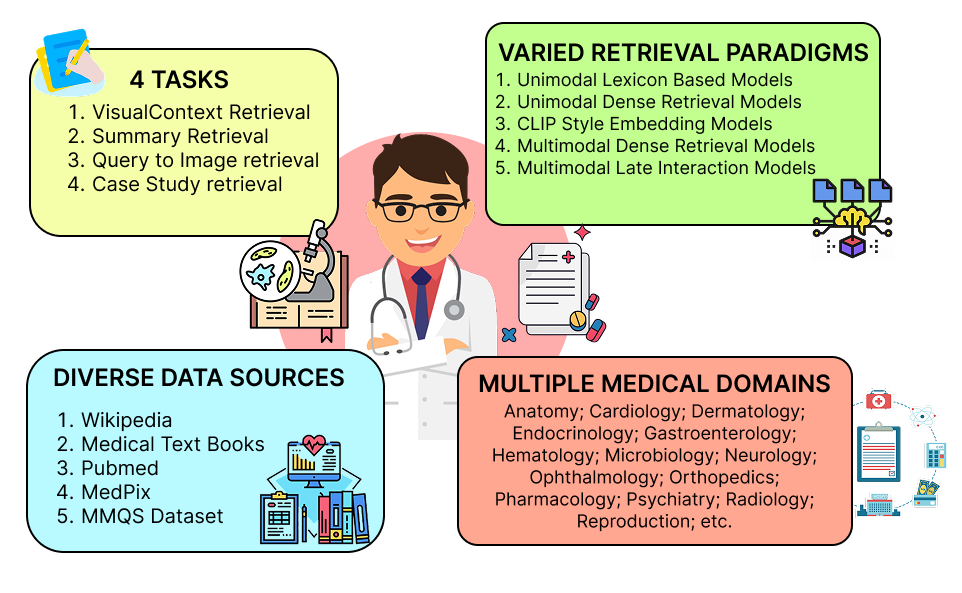}
    \caption{ \textit{M3Retreive} is a multimodal medical retrieval benchmark comprising samples from \textbf{four} different tasks
 across \textbf{multiple healthcare subdomains} obtained from a variety of open-sourced data sources resulting in total dataset size of about 800K query-corpus pairs. It encompasses testing of retrieval models across varied retrieval paradigms}
    \label{fig:hero_image}
\end{figure*}
Retrieval models play a crucial role in efficiently accessing and utilizing the vast amounts of information available today. These models facilitate the quick and accurate extraction of relevant data,
essential for informed decision-making in various downstream applications across numerous domains. With the emergence of Retrieval-Augmented Generation (RAG) \citep{lewis2021retrievalaugmentedgenerationknowledgeintensivenlp}, the importance of high-quality retrieval systems has grown exponentially, particularly in knowledge-intensive domains. 

In recent years, advancements in deep learning have facilitated the development of multimodal retrieval models that process and generate embeddings from both textual and visual data. This capability is particularly significant in the medical domain, where images such as X-rays, MRIs, and histopathological slides provide critical context alongside textual descriptions, a fact that has already been highlighted in works like LLaVa-Med \citep{li2023llavamedtraininglargelanguageandvision} and MedSumm \cite{ghosh2024medsumm}. The importance of both image and text embedding for effective knowledge extraction for the Medical domain has been highlighted in existing works like  \citep{ghosh2024clipsyntel,ghosh2024healthalignsumm,ghosh2024sights}. The performance of multimodal retrievers is vital for downstream tasks such as multimodal information extraction \citep{sun2024umie}, question answering \citep{luo2023unifying}, and cross-modal retrieval \citep{wang2025cross}, as it directly impacts the accuracy of generated content. This becomes especially critical in safety-sensitive domains like healthcare, where trust and reliability are paramount. However, despite these developments, there is currently no standardized benchmark to evaluate the performance of multimodal retrieval models in medical applications.

\textbf{Research Gap:} Though efforts have been made to develop extensive retrieval benchmarks, including the BEIR Benchmark \citep{thakur2021beirheterogenousbenchmarkzeroshot} and the M-BEIR Benchmark \citep{wei2023uniirtrainingbenchmarkinguniversal} for the evaluation of text-only retrievals and multimodal retrievals,  respectively, their expansion into domain-specific tasks, such as medical retrieval, remains an open challenge.  Besides being an important domain of study for NLP and machine learning applications in general, the importance of benchmarking in this domain cannot be overstated. The medical field is highly complex, and access to precise and relevant information can significantly influence patient care and medical research. An intricate field like Medicine poses a unique set of challenges, such as: a) \textbf{Complex Medical Terminologies:} The medical field uses complex, variable terminology that often requires plain-language explanations for clarity; for instance, \textit{Acute Hemorrhagic Leukoencephalitis}—a severe brain inflammation—may be described as causing \textit{a sudden, severe headache and episodes of confusion}\footnote{\url{https://www.medicalnewstoday.com/articles/acute-hemorrhagic-leukoencephalopathy}}, highlighting the need for retrieval systems to accurately interpret such terms. b) \textbf{Multiple niche specialities:} Medicine is divided into many specialized disciplines, each requiring tailored methods to address specific patient needs\footnote{\url{https://en.wikipedia.org/wiki/Medical_specialty}}\footnote{\url{https://www.abms.org/wp-content/uploads/2021/12/ABMS-Guide-to-Medical-Specialties-2022.pdf}}; as hospitals are organized by specialty, retrieval systems must be finely evaluated to assess their generalization across diverse medical domains. c) \textbf{Complex Image-Text Relationship:} Medical images can appear similar yet represent different conditions when combined with patient history; for example, \textit{Viral Exanthems} and \textit{Drug Eruptions} are hard to distinguish without detailed context\footnote{\url{https://www.slideshare.net/slideshow/viral-exanthemsmodule/38060053}}, requiring multimodal retrieval systems to jointly encode and interpret image-text data accurately.\par
Existing medical datasets in the BEIR Benchmark focus on a  \textbf{single (textual) modality} and lack the necessary scale and diversity required for fine-grained assessment of retrieval performance across \textbf{multiple medical disciplines}. Medical text-image pair datasets have not been explicitly covered in any of the datasets included in the M-BEIR Benchmark, thus creating a pressing need for a comprehensive benchmark tailored to the medical domain that evaluates retrieval models on real-world, multimodal data.
Table \ref{tab:compact_medical_retrieval} provides a comprehensive overview of existing multimodal medical datasets for retrieval, highlighting how \name distinguishes itself in terms of task complexity and the broad range of domains it covers.

\begin{table*}[htbp]
  \centering
  \scriptsize
  \setlength{\tabcolsep}{6pt}
  \renewcommand{\arraystretch}{1.1}
  \caption{Comparison of Benchmark Retrieval Datasets in the Medical Domain}
  \label{tab:compact_medical_retrieval}
  \begin{tabularx}{\textwidth}{@{} L  c  L  L  c  c @{}}
    \toprule[1.2pt]
    \rowcolor{blue!25}
    \textbf{Benchmark} & \textbf{Data Points} 
      & \textbf{Task} 
      & \textbf{Medical Domains} 
      & \textbf{Modalities} 
      & \textbf{Open?} \\
    \midrule
    NFCorpus~\cite{boteva2016full}
      & 3,244 queries, 9,964 docs
      & Text $\to$ Text
      & Nutrition; General Medicine
      & Docs; Queries
      & Yes \\

    TREC-COVID~\cite{voorhees2021trec}
      & 171,332 articles
      & Text $\to$ Text
      & COVID-19 Research
      & Articles; Queries
      & Yes \\

    MIMIC-CXR~\cite{johnson2019mimic}
      & 377,110 X-rays; 227,835 reports
      & Text $\leftrightarrow$ Image
      & Chest Radiology
      & X-rays; Reports
      & Yes \\

    ImageCLEFmed~\cite{ImageCLEFmedicalCaption2024}
      & $\sim$66,000 images
      & Multi-query (Text/Image $\to$ Images)
      & Radiology; Pathology; Dermatology
      & X-ray, CT, MRI; Reports
      & Yes \\

    3D-MIR~\cite{abacha20233dmirbenchmarkempiricalstudy}
      & 4 anatomies (Colon, Liver, Lung, Pancreas)
      & 3D CT $\to$ Volume
      & Multi-organ Imaging
      & 3D CT
      & Yes \\

    BIMCV-R~\cite{chen2024bimcv}
      & 8,069 CT volumes
      & Text $\to$ Image
      & Respiratory (COVID; Pneumonia)
      & CT; Reports
      & Yes \\

    CBIR (TotalSegmentator)~\cite{li2021recent}
      & 29 coarse; 104 detailed regions
      & Region-based (Segment $\to$ Scan)
      & Multi-organ Anatomy
      & Volumetric Scans
      & No \\
    \addlinespace

    \textbf{{\em M3Retrive} (Ours)}
      & $\sim$ 164947 queries; $\sim$ 1238038 docs
      & Multimodal (Text+Image $\to$ Text/Image)
      & 16 specialties\newline(e.g., Anatomy, Cardio., Pulmo., Derm., Endo., Neuro., Radiol.)
      & Docs; Queries
      & Yes \\
    \bottomrule[1.2pt]
  \end{tabularx}

  \vspace{1ex}
  \begin{flushleft}
    \footnotesize
    Abbreviations: CT = computed tomography; MRI = magnetic resonance imaging.
  \end{flushleft}
\end{table*}

\textbf{Present Work:} We introduce \name , a Multimodal Medical Retrieval Benchmark designed to bridge the gap in medical information retrieval. By integrating both textual and visual data, \textbf{M3Retrieve} enables a more realistic evaluation of retrieval models in complex multimodal medical contexts.
The key contributions of this work can be summarised as :\par
\vspace{-0.2cm}
\textbf{a) Introduction of  \name.} We present the first large-scale multimodal retrieval benchmark for the medical domain. \textbf{\textit{M3Retrieve}} accepts multimodal queries and targets realistic document stores spanning multiple specialties.

b) \textbf{Comprehensive Dataset.}  \textbf{\textit{M3Retrieve}}  aggregates 22 manually-curated datasets (all under permissive licences) that cover \textbf{16 medical disciplines} and comprise \textbf{920\,K text documents} plus \textbf{818\,K multimodal queries}, providing broad coverage of real-world clinical scenarios.\par

\textbf{c) Clinically-Grounded Task Suite.} Guided by consultations with healthcare professionals, we define five retrieval tasks that mirror routine information-seeking workflows: \textbf{Visual Context Retrieval} (image + short text/caption $\rightarrow$ relevant passage), \textbf{Multimodal Query-to-Image Retrieval} (image or text description $\rightarrow$ visually similar image), \textbf{Case Study Retrieval} (image + patient transcript $\rightarrow$ closest full past case), and \textbf{Multimodal Summarisation Retrieval} (long report + associated images $\rightarrow$ concise summary). \par
\textbf{d) Systematic Performance Evaluation} We benchmark several state-of-the-art multimodal retrieval models on \name, revealing discipline-specific challenges and quantifying their impact on retrieval effectiveness.\par


\section{Related Works}
\subsection{Retrieval Benchmarks}
 The evaluation of retrieval systems has a long and rich history. Early efforts, rooted in the Cranfield paradigm and later formalized by initiatives such as the Text Retrieval Conference (TREC) \citep{wiki:TextRetrievalConference}, established core evaluation measures such as precision, recall, and mean average precision—that continue to underpin retrieval performance assessment today. Over time, large-scale benchmarks like MS MARCO \citep{bajaj2018msmarcohumangenerated} and BEIR \citep{thakur2021beirheterogenousbenchmarkzeroshot} have provided standardized test collections and protocols for open-domain text retrieval, thereby driving the development of more robust retrieval models. In the medical domain, the unique nature of clinical language and the critical need for factual correctness have spurred the creation of specialized benchmarks. Initiatives such as BioASQ \citep{jeong2021transferabilitynaturallanguageinference}, PubMedQA \citep{jin2019pubmedqa}, and other medical question answering datasets \citep{ngo2024comprehensivepracticalevaluationretrievalaugmented} have primarily focused on text retrieval and QA tasks, evaluating models on their ability to retrieve and reason over biomedical literature. However, these benchmarks rarely incorporate non‐textual data even though medical diagnosis and decision support often require the interpretation of images (e.g., radiographs or histology slides) alongside text. Language-specific and domain-specific retrieval benchmarks have further refined evaluation criteria by addressing nuances in linguistic usage. For instance, initiatives like MIRACL \citep{zhang2022makingmiraclmultilingualinformation}, mMARCO \citep{bonifacio2022mmarcomultilingualversionms}, and various language-specific benchmarks \citep{acharya2024benchmarkingbuildingzeroshothindi, snegirev2025russianfocusedembeddersexplorationrumteb} have motivated the development of more effective multilingual retrieval models. Similarly, fine-grained domain-specific analyses in BEIR-like benchmarks have fostered the advancement of domain-agnostic embedding models. The most prominent and extensive multimodal retrieval benchmark to date is UniIR \citep{wei2023uniirtrainingbenchmarkinguniversal}. However, the datasets within UniIR do not comprehensively address the intricate challenges of the medical domain. Medical applications demand fine-grained clinical detail, domain-specific terminologies, and the integration of both visual and textual evidence to support accurate decision-making and thus demand a separate benchmark of their own for a more holistic evaluation.

\subsection{Retrieval Models}
The evolution of retrieval systems in Natural Language Processing (NLP) has progressed from lexicon-based models to advanced dense and multimodal architectures \cite{ghosh2024exploring}. Early retrieval models like BM-25 \citep{10.1561/1500000019} relied on lexicons. With the rise of deep learning, dense retrieval models, such as E5 \citep{wang2022text}, BGE \citep{bge_embedding}, and NV Embed \citep{lee2025nvembedimprovedtechniquestraining}, have addressed vocabulary mismatch by utilizing contrastive learning to produce semantic embeddings. The integration of multiple modalities began with CLIP \citep{radford2021learningtransferablevisualmodels}, aligning visual and textual representations in a shared embedding space for cross-modal retrieval. This approach was extended in models like MM Ret \citep{zhou2024megapairsmassivedatasynthesis} and MedImageInsight \citep{codella2024medimageinsightopensourceembeddingmodel}, which specialize in medical image-text retrieval. Unified multimodal retrievers such as the UniIR family \citep{wei2023uniirtrainingbenchmarkinguniversal}( CLIP SF and BLIP FF) enable cross-modal retrieval across diverse data types. Recent models like VLM2Vec \citep{jiang2025vlm2vectrainingvisionlanguagemodels} and MM Embed \citep{lin2024mmembeduniversalmultimodalretrieval} further improve joint representation learning for text and images. Additionally, late-interaction models like FLMR \citep{lin2023finegrainedlateinteractionmultimodalretrieval} compute token-level similarities for more precise retrieval relevance determination. 
We believe that with the introduction of \textit{M3Retrieve}, the community will be better equipped to assess discipline-specific challenges and the integration of multimodal signals, thereby helping in establishing practical guidelines for developing more reliable medical retrieval systems.

We present \name , a comprehensive multimodal medical retrieval benchmark for healthcare domain that consists of 5 different tasks. In this section, we provide a detailed formulation for each of the tasks decided, outline the diverse data sources utilized in constructing the  \name , and describe the methodology employed to curate the benchmark from these sources.\par
\section{Creation of the {\em M3Retrieve} Benchmark}
The following section outlines the tasks included in the \textit{M3Retrieve} Benchmark and describes the methodology used for their construction. Table \ref{tab:dataset_stat} gives an overview of the number of corpus and queries for each of the tasks in the \textit{M3Retrieve} Benchmark.

\begin{table}[]
    \centering
    \small
    \begin{tabular}{l|rr}
    \toprule
         \textbf{Task} & \textbf{\#Corpus} & \textbf{\# Queries} \\
         \midrule
         Visual Context Retrieval& 507101 & 93488 \\
         Summary Retrieval & 228887 & 3015\\
         Query to Image Retrieval & 2050 & 671\\
         Case study Retrieval & 500000 & 67773 \\ \hline
       
    \textbf{Total} & 1238038 & 164947
\\ 
\bottomrule
    \end{tabular}
    \caption{Statistics of the Dataset in the \textit{M3Retrieve} Benchmark showing the number of corpus and query in the evaluation set for each task in the benchmark.}
    \label{tab:dataset_stat}
\end{table}
\subsection{VisualContext Retrieval}

\begin{figure}
    \centering
    \includegraphics[width=1\linewidth]{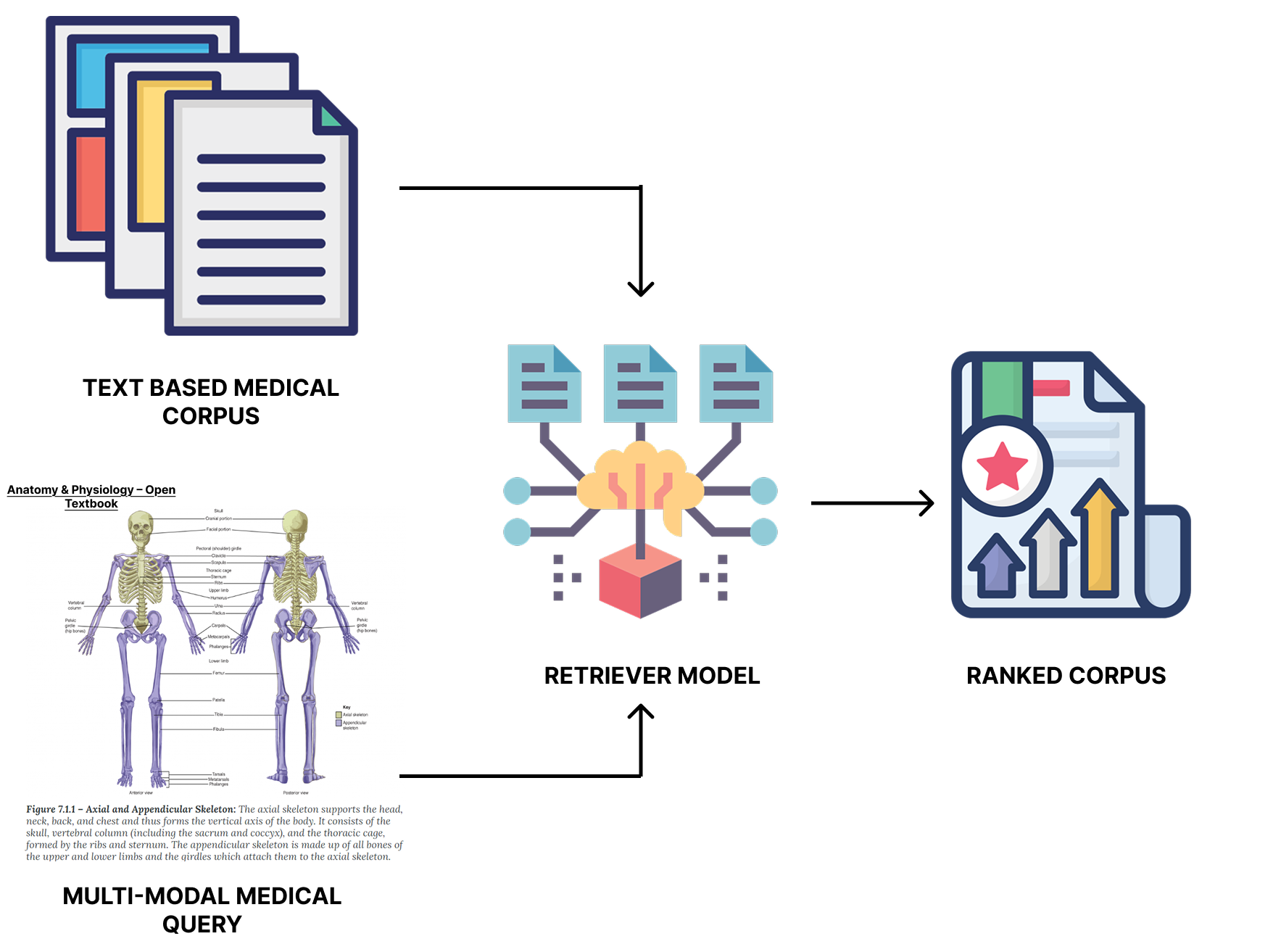}
    \caption{Overview of a retrieval task addressed in the \textit{M3Retrieve} Benchmark. The task aims to integrate both text and image data, with the retriever model ranking documents based on relevance. The multimodal framework enriches retrieval performance by incorporating visual information alongside traditional text-based retrieval.}
    \label{fig:task_def}
\end{figure}

\textbf{\textit{Task Formulation:}} Given a multimodal query \( Q = (Q_{\text{text}}, Q_{\text{image}}) \) and a retrieval corpus \( D = \{D_1, D_2, \dots, D_n\} \), the retriever aims to find and rank the set of relevant documents:\[
D_{Q}^{+} = \{D_{Q,1}^{+}, \dots, D_{Q,m}^{+}\} \subset D, \quad \text{where } m \ll n
\]

where,  \textbf{Positive documents \(D_{Q}^{+}\)} represents directly relevant documents containing information aligned with the query's text and image. 
\subsubsection{Data Curation and Relevance Design}

\textbf{Data Sources}:
For the VisualContext Retrieval task we collect data from three primary open-access sources namely: \textbf{(i)} Wikipedia, using the Wikimedia API\footnote{\url{https://api.wikimedia.org/wiki/Getting_started_with_Wikimedia_APIs}} to extract structured multimodal content across diverse medical domains, yielding 288,983 corpus documents and 24,523 multimodal queries; \textbf{(ii)} PubMed, leveraging a subset of 10,000 articles from the dataset by Li et al. \cite{li2023llavamedtraininglargelanguageandvision}, resulting in 204,217 corpus documents and 64,366 text-image queries; and \textbf{(iii)} Open Access medical textbooks sourced from Creative Commons licensed platforms such as OpenStax and Open Oregon State, processed using similar extraction and pairing strategies.

\textbf{Relevancy Mapping Formulation}:
Each data source presented unique challenges, so we collaborated with medical experts to define dataset-specific relevancy mappings. For Wikipedia, experts noted that images are generally relevant to the entire article, making it difficult to distinguish between related and unrelated paragraphs. Consequently, for Wikipedia, we defined the relevance mapping as follows:

\begin{tcolorbox}[title=\textbf{Relevancy Mapping (Wikipedia)}, colback=green!5!white, colframe=green!75!black]
\textbf{Strategy:} Images explicitly referred to in a paragraph were assigned a relevance score of \textbf{2}. All other paragraphs within the same article received a score of \textbf{1} due to their contextual relation to the overall topic.
\end{tcolorbox}

We used 10,000 PubMed articles from \cite{li2023llavamedtraininglargelanguageandvision}  and, based on expert input, assigned the highest relevance to the paragraph referencing a figure, as its significance diminishes elsewhere

\begin{tcolorbox}[title=\textbf{Relevancy Mapping (PubMed)}, colback=green!5!white, colframe=green!75!black]
\textbf{Strategy:} The paragraph that explicitly mentions the figure was assigned a score of \textbf{2}, while all other paragraphs were excluded from consideration. This reflects the highly localized relevance of figures in scientific articles.
\end{tcolorbox}

In textbooks, each image-caption pair forms a multimodal query, with paragraphs as candidate documents; figures are typically relevant only within their specific textbook section.

\begin{tcolorbox}[title=\textbf{Relevancy Mapping(TextBooks)}, colback=green!5!white, colframe=green!85!black]
\textbf{Strategy:} The paragraph that directly references a figure was assigned a score of \textbf{2}. All other paragraphs within the same section were considered contextually relevant and given a score of \textbf{1}.
\end{tcolorbox}

This unified scoring framework across data sources allowed the creation of a consistent, high-quality benchmark for evaluating multimodal evidence retrieval systems.

\subsection{Multimodal Summary Retrieval}

\paragraph{\textbf{\textit{Task Formulation.}}}
Given a multimodal context  
\(Q = (Q_{\text{text}},\, Q_{\text{image}})\),  
where \(Q_{\text{text}}\) may represent a clinical note, patient conversation, or textual report, and \(Q_{\text{image}}\) is an associated medical image (e.g., X-ray, MRI),  
the goal is to retrieve the most relevant summary from a candidate summary pool  
\[
S = \{S_1, S_2, \dots, S_n\},
\]  
where each \(S_i\) is a standalone summary. Exactly one candidate \(S_Q^{\star}\) is considered the correct summary, which best captures the key information from both modalities.

\subsubsection{Data Curation and Relevance Design}
\textbf{Data Sources.} We use the MMQS dataset \cite{ghosh2024clipsyntel} for the task of multimodal healthcare summarization. MMQS consists of 3,015 curated samples, where each instance comprises a multimodal medical query—combining a textual component (e.g., patient query or clinical dialogue) and a visual component (e.g., a relevant medical image)—paired with a corresponding summary that integrates information from both modalities. The summaries are drawn from a larger retrieval corpus derived from HealthcareMagic \cite{mrini2021joint}, which contains 228,887 medical queries and their corresponding expert-written summaries. This corpus serves as the foundation for candidate summaries during retrieval, providing a rich pool of medical knowledge spanning both textual and visual contexts. Here, we augment the original HealthcareMagic dataset \cite{mrini2021joint} by appending the curated multimodal summaries from MMQS, thereby constructing an expanded summarization corpus for retrieval.\par
\textbf{Relevancy Mapping Formulation}

Each multimodal medical query serves as the input, and the retriever is tasked with identifying the corresponding summary from a large-scale summarization corpus derived from HealthcareMagic \cite{mrini2021joint}.

\begin{tcolorbox}[title=\textbf{Relevancy Mapping}, colback=green!5!white, colframe=green!75!black]
\textbf{Strategy:} The summary that directly corresponds to the given multimodal query is assigned a score of 2, while all other summaries in the corpus are assigned a score of 0.
\end{tcolorbox}

\subsection{Multimodal Query to Image Retrival }

\paragraph{\textbf{\textit{Task Formulation.}}}
Given a multimodal query  
\(Q = (Q_{\text{text}},\, Q_{\text{image}})\),  
where \(Q_{\text{text}}\) represents a textual medical query or dialogue and \(Q_{\text{image}}\) provides visual context (e.g., an indicative or reference image),  
the task is to retrieve the most relevant image from a candidate image pool  

\[
I = \{I_1, I_2, \dots, I_n\},
\]  
where each \(I_i\) is a standalone medical image (e.g., chest X-ray, MRI, ultrasound, pathology scan).  
Exactly one candidate image \(I_Q^{\star}\) is considered the correct or best matching image for the query. 

\subsubsection{Data Curation and Relevance}

\textbf{Data Sources}: To support the novel retrieval task of selecting relevant medical images based on a multimodal input query (consisting of textual and visual cues), we curated a dataset derived from the public MedPix 2.0 \cite{siragusa2024medpix} repository—a comprehensive radiology teaching file provided by the U.S. National Library of Medicine.  MedPix 2.0 is the best dataset for this task because it offers expertly curated, semantically rich text–image pairs across diverse medical conditions, enabling precise and clinically grounded multimodal query-to-image relevance mapping.

\textbf{Relevancy Mapping Formulation:} To establish reliable ground-truth for the multimodal image query $\rightarrow$ image retrieval task, we leverage the structural integrity of the MedPix 2.0 dataset, where each clinical case is identified by a unique \texttt{U\_id}. All images associated with the same \texttt{U\_id} are considered \textit{relevant} to the query composed from that case's textual description and accompanying visual clue.

\begin{tcolorbox}[title=\textbf{Relevancy Mapping}, colback=green!5!white, colframe=green!75!black]
\textbf{Strategy:} All images that share the same \texttt{U\_id} as the multimodal query (i.e., derived from the same clinical case) are assigned a score of 2 and all other images in the corpus are assigned a score of 0.
\end{tcolorbox}

\subsection{Case Study Retrieval}

\paragraph{\textbf{\textit{Task Formulation.}}}
Given a multimodal clinical query  
\(Q = (Q_{\text{text}},\, Q_{\text{image}})\),  
where \(Q_{\text{text}}\) represents a patient complaint, diagnostic note, or medical dialogue, and \(Q_{\text{image}}\) is an associated clinical image (e.g., scan, X-ray, pathology slide),  
the goal is to retrieve the most relevant case study from a set of documented medical cases

\[
S = \{S_1, S_2, \dots, S_n\},
\]  
where each \(S_i\) is a structured medical case study consisting of textual findings, diagnoses, and possibly images.  
Exactly one case study \(S_Q^{\star}\) is most relevant to the query.

\subsubsection{Data Curation and Relevance}

\textbf{Data Sources.} To enable the retrieval of clinically relevant case studies based on multimodal (Image + Textual Description), we make use of the MultiCaRe \cite{multicare2024} dataset—a publicly available resource constructed from open-access case reports on PubMed Central. This dataset offers a comprehensive collection of over 93,000 de-identified clinical cases paired with more than 130,000 diagnostic images. Its broad medical coverage and well-aligned text–image pairs make it an ideal foundation for developing and evaluating multimodal retrieval systems in real-world healthcare settings.

\textbf{Relevance Mapping Formulation}
To establish reliable ground-truth for the multimodal relevance mapping task, we leverage the structured design of the MultiCaRe dataset, where each clinical case is tagged by a unique case\_id. All textual narratives and associated images sharing the same case\_id are considered relevant to any query derived from that case’s combined textual and visual information.

\begin{tcolorbox}[title=\textbf{Relevancy Mapping}, colback=green!5!white, colframe=green!75!black]
\textbf{Strategy:} All case studies that share the same \texttt{case\_id} as the multimodal query (i.e., originate from the same clinical record) are assigned a score of 2. Case studies with similar diagnostic categories or overlapping symptoms but from different records are assigned a score of 1, while all remaining case studies in the corpus are assigned a score of 0.
\end{tcolorbox}

\subsection{Quality Control Using Domain Expert }
Throughout the data curation process, medical experts provided valuable feedback to ensure the selection of the most relevant data sources and essential medical modalities, enhancing the dataset’s quality and applicability. Their insights were instrumental in establishing accurate relevance mappings, ensuring that query-document relationships aligned with real-world medical reasoning. Additionally, to validate the dataset’s reliability, a sample of 80 queries across each task  was reviewed by two doctors. The evaluation yielded a \emph{Cohen’s kappa score of 0.78}, indicating a high level of agreement between the reviewers and confirming that the dataset rankings were accurate and meaningful.

\section{Experimental Setup }
All experiments were conducted using the MTEB Python library \footnote{\url{https://github.com/embeddings-benchmark/mteb}} on NVIDIA A100 80GB GPUs. FLMR was evaluated using the implementation available at \url{https://github.com/LinWeizheDragon/FLMR}, and document retrieval was performed using BM-25 via Pyserini \footnote{\url{https://github.com/castorini/pyserini}}. For both FLMR and BM-25, the evaluation metrics were computed using the pytrec\_eval \footnote{\url{https://github.com/cvangysel/pytrec_eval}} Python library, following the implementation in the MTEB library. We used nNDCG@10 as the primary metric for evaluation.

\subsection{Baseline Models}

To evaluate retrieval performance on the \name Benchmark, we assess a range of uni-modal and multimodal models, categorized as follows:

\begin{itemize}
    \item \textbf{Lexicon-Based Model:} \textbf{BM25} \cite{10.1561/1500000019} — A strong traditional baseline using term frequency and inverse document frequency for scoring.

    \item \textbf{Text-Based Encoders:} \textbf{E5-Large-v2} \cite{wang2022text} (1024-dim; weakly-supervised contrastive learning), \textbf{BGE-en-Large} \cite{bge_embedding} (1024-dim; top MTEB performance), \textbf{NV-Embed-v2} \cite{lee2025nvembedimprovedtechniquestraining} (4096-dim; MTEB leader with latent-attention pooling).

    \item \textbf{CLIP-Style Models:} \textbf{MMRet-Large} \cite{zhou2024megapairsmassivedatasynthesis} (CLIP-based; 768-dim; context length 77), \textbf{MedImageInsight (MII)} \cite{codella2024medimageinsightopensourceembeddingmodel} (medical domain; CLIP-style contrastive learning), \textbf{CLIP-SF} \cite{wei2023uniirtrainingbenchmarkinguniversal} (768-dim; context length 77).

    \item \textbf{Multimodal Encoders:} \textbf{BLIP-FF} \cite{wei2023uniirtrainingbenchmarkinguniversal} (BLIP-based; 768-dim; context length 512), \textbf{MM-Embed} \cite{lin2024mmembeduniversalmultimodalretrieval} (extends NV-Embed-v1; state-of-the-art on UniIR and MTEB).

    \item \textbf{Multimodal Late Interaction Retriever:} \textbf{FLMR} \cite{lin2023finegrainedlateinteractionmultimodalretrieval} — Uses token-level similarity for fine-grained late interaction between queries and documents.
\end{itemize}

\vspace{-0.5cm}

\section{Results Analysis}


\begin{table*}[]
    \centering
    \small
    \begin{tabular}{l|cccc}
\toprule
\textbf{Method} & \textbf{VisualContext Retrieval} & \textbf{Summary Retrieval}& \textbf{Query to Image Retrieval} & \textbf{Case Study Retrieval} \\
\midrule
\ccrrow BM-25 & 38.07 & 18.16 & N/A & \textbf{11.50} \\
\midrule
\ccbrow E5 Large & 35.14 & 70.23 & N/A & 7.68 \\
\ccbrow BGE & 32.32 & 83.66 & N/A & 6.59 \\
\ccbrow NV Embed & \underline{43.28} & \textbf{89.73} & N/A & \underline{10.99} \\
\midrule
\ccgrow MM Ret & 24.56 & 43.71 & 2.27 & 1.09 \\
\ccgrow MII & 28.13 & 22.5 & \textbf{43.53} & 1.64 \\
\ccgrow CLIP SF & 26.44 & 26.30 & 29.06 & 1.27 \\
\midrule
\ccyrow BLIP FF & 24.72 & 20.89 & 2.23 & 0.92 \\
\ccyrow MM Embed & \textbf{45.47} & \underline{76.27} & \underline{29.49} & 9.91 \\
\midrule
\ccmrow FLMR & 24.80 & 21.30 & 2.56  & 1.48 \\
\bottomrule
    \end{tabular}
     \caption{NDCG@10 scores for ten retrieval models representing different retrieval styles, including \ccr{lexicon-based retrievers}, \ccb{uni-modal dense retrievers}, \ccg{CLIP-style} models, \ccy{multi-modal dense retrievers}, and \ccm{late-interaction multi-modal retrievers} on the \name Benchmark. The best-performing model has been highlighted as \textbf{bold} while the second-best model has been \underline{underlined}.}
    \label{tab:resuts}
\end{table*}

To assess the effectiveness of various retrieval models in the medical domain, we evaluated multiple uni-modal and multimodal approaches on the \name . The models analyzed can be divided into five broad categories: \textbf{lexicon-based}, \textbf{text-based dense encoders}, \textbf{CLIP-style multimodal retrievers}, \textbf{multimodal encoders}, and \textbf{late-interaction models}.

Table~\ref{tab:resuts} presents the NDCG@10 scores for various retrieval models across the four tasks in the \name Benchmark. The models are categorized into five retrieval paradigms: lexicon-based retrievers, uni-modal dense retrievers, CLIP-style models, multi-modal dense retrievers, and late-interaction multi-modal retrievers.

It should also be noted that the reported results are grounded in the structural assumptions of the underlying datasets. 

\subsection{Overall Performance Trends}

Multimodal models exhibit significant potential, particularly in tasks that inherently require the integration of textual and visual information. For instance, in the \textbf{VisualContext Retrieval} task, the {\bf MM-Embed} model achieves the highest NDCG@10 score of 45.47, outperforming all other models. Similarly, in the \textbf{Query to Image Retrieval} task, the {\bf MedImageInsight} model leads with a score of 43.53. These results underscore the advantage of multimodal models in scenarios where both text and image modalities are crucial.

However, in tasks that are predominantly textual, such as \textbf{Summary Retrieval} and \textbf{Case Study Retrieval}, uni-modal dense retrievers demonstrate superior performance. The {\bf NV-Embed} model achieves the highest scores in both tasks, with 89.73 and 10.99, respectively. This suggests that, in the current landscape, uni-modal models remain highly effective for text-centric retrieval tasks.

\subsection{Task-Specific Model Performance}

\paragraph{VisualContext Retrieval:} This task benefits from models capable of integrating multimodal information. The {\bf MM-Embed} model achieves the highest performance (45.47), followed by the {\bf NV-Embed} (43.28) and {\bf BM25} (38.07). The strong performance of {\bf MM-Embed} highlights the effectiveness of multimodal dense retrievers in capturing the nuanced relationships between text and images.

\paragraph{Summary Retrieval:} Uni-modal dense retrievers dominate this task, with {\bf NV-Embed} achieving the top score of 89.73, followed by {\bf BGE} (83.66) and {\bf E5 Large} (70.23). Multimodal models lag behind, indicating that current multimodal approaches may not yet effectively handle tasks that are primarily textual.

\paragraph{Query to Image Retrieval:} The {\bf MedImageInsight} model leads with a score of 43.53, demonstrating the strength of CLIP-style models in image retrieval tasks. The {\bf MM-Embed} model follows with 29.49, and {\bf CLIP SF} achieves 29.06. These results suggest that models trained with contrastive learning on image-text pairs are particularly effective for image-centric retrieval tasks.

\paragraph{Case Study Retrieval:} The {\bf NV-Embed} model again achieves the highest score (10.99), indicating that uni-modal dense retrievers are currently more effective for retrieving comprehensive case studies. The {\bf MM Ret} model follows closely with 10.87, suggesting some potential for multimodal models in this area.

\section{Conclusion}
We introduce \name, the first comprehensive multimodal Medical Retrieval Benchmark, designed to evaluate retrieval models across 16+ medical domains, covering 500K+ documents and 100K queries. This benchmark rigorously assesses uni-modal (text-based) and multimodal retrieval models across five retrieval styles and architectural approaches, providing a detailed analysis of their performance in complex medical scenarios. Given the critical role of accurate medical information retrieval in patient care, clinical decision-making, and research, this benchmark helps identify gaps and strengths in current models. By benchmarking a diverse range of retrieval approaches, \name establishes a strong foundation for developing more effective and specialized multimodal retrieval systems. We believe it will be instrumental in driving future research and advancing medical AI systems to enhance real-world healthcare applications.

\section {Ethical Considerations}
All data in \name comes from publicly available sources with Creative Commons (CC) licenses, ensuring compliance with HIPAA and GDPR. Medical professionals were involved throughout the dataset design to ensure relevance and accuracy. A human evaluation was conducted post-construction to verify quality and fairness, reinforcing ethical and responsible medical benchmark.

\section{Limitation and Future Work}
The \name is a foundational step toward a diverse multi-domain, multimodal medical retrieval benchmark, but it has certain limitations. Currently, it covers 16 broad medical domains, which may not encompass all specialties. Future work will expand coverage to additional disciplines. The benchmark currently features multimodal queries with text-only documents, a common setup, but future versions will incorporate more modalities for both queries and corpora. Our findings show that existing models underperform in medical retrieval compared to general-domain tasks, underscoring the need for a medical-specific multimodal retrieval model. 
Additionally, our figure-to-paragraph mappings were validated only through a limited manual review, and large-scale verification across the full dataset was not feasible, making our approach reliant on the assumption that figures are responsibly referenced in the original publications.

\section{Acknowledgement}
We thank the ACL ARR reviewers for their constructive comments and valuable feedback on the draft. Akash Ghosh and Sriparna Saha extend their sincere appreciation to the SERB (Science and Engineering Research Board) POWER scheme, Department of Science and Engineering, Government of India of India, for generously funding this research.

\bibliography{latex/acl_latex}

\appendix

\section{Appendix}
\label{sec:appendix}
\subsection{Example of datapoints in the \textit{M3Retrieve Benchmark}}
\begin{figure*}
    \centering
    \includegraphics[width=1\linewidth,height=4cm]{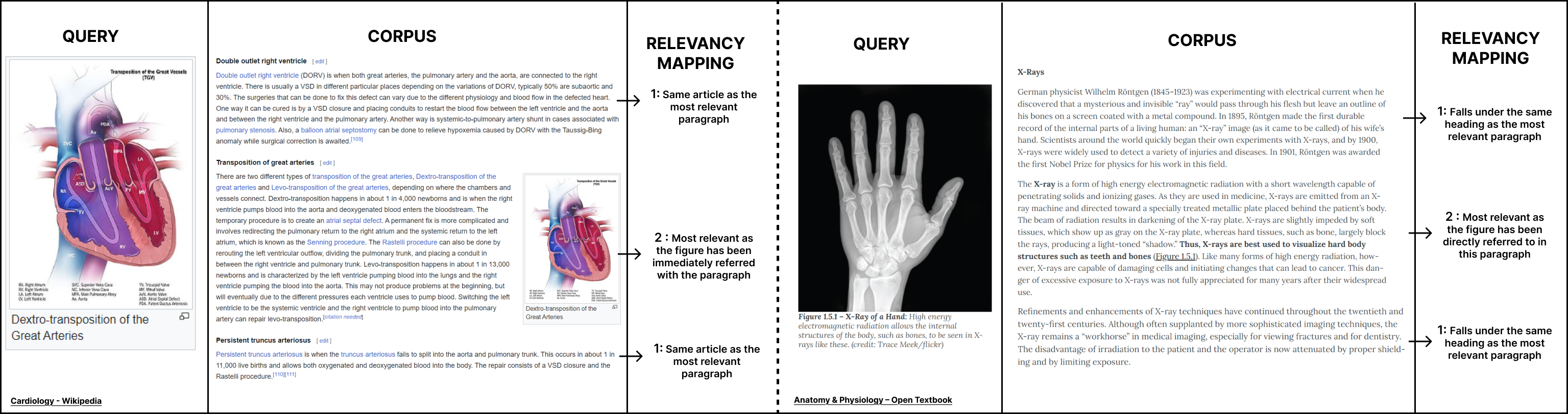}
    \caption{An example from the \textit{M3Retrieve} Benchmark showing the query image-text pair and corpus texts along with justifications for the assigned scores.}
    \label{fig:enter-label}
\end{figure*}

Figure \ref{fig:enter-label} exhibits an example from the \textbf{\textit{M3Retrieve}} Benchmark showing the query image-text pair and corpus texts along with justifications for the assigned scores.

\subsection{Discipline wise analysis of the tasks in the \textit{M3Retrieve} Benchmark}
\subsubsection{Visual Context Retrieval}
\begin{table*}[!htb]
    \centering
    \tiny

\begin{tabular}{l|%
\ccrcol|%
\ccbcol\ccbcol\ccbcol|%
\ccgcol\ccgcol\ccgcol|%
\ccycol\ccycol|%
\ccmcol}                         
\toprule
Discipline & BM-25 & E5 & BGE & NV Embed & MM Ret & MII & CLIP SF & BLIP FF & MM Embed & FLMR \\
\midrule
Anatomy and Physiology & 61.20 & 49.17 & 60.15 & \textbf{66.29} & 47.25 & 49.15 & 47.96 & 52.91 & \underline{64.14} & 2.05 \\
Cardiology & 39.25 & 33.26 & 31.15 & \underline{42.16} & 26.12 & 27.56 & 27.66 & 26.32 & \textbf{45.3} & 32.02 \\
Dermatology & 41.78 & 42.26 & 34.10 & \underline{51.53} & 33.37 & 36.27 & 34.70 & 33.49 & \textbf{52.59} & 35.05 \\
Endocrinology and Diabetes & 35.80 & 36.99 & 33.86 & \underline{45.83} & 28.95 & 29.83 & 27.97 & 30.45 & \textbf{46.38} & 32.86 \\
Gastroenterology & 35.16 & 36.90 & 31.67 & \textbf{46.07} & 26.47 & 29.41 & 27.96 & 27.68 & \underline{46.01} & 33.20 \\
Hematology & 32.23 & 31.89 & 29.00 & \underline{38.46} & 22.65 & 24.01 & 23.65 & 20.53 & \textbf{39.73} & 27.47 \\
Microbiology and Cell Biology & 34.01 & 30.65 & 29.23 & \underline{39.92} & 19.14 & 19.35 & 18.64 & 19.14 & \textbf{39.80} & 6.55 \\
Miscellaneous & 34.25 & 31.29 & 29.13 & \underline{39.76} & 21.05 & 21.12 & 21.07 & 20.89 & \textbf{41.93} & 25.69 \\
Neurology and Neuroscience & 24.54 & 23.15 & 19.88 & \underline{28.04} & 15.94 & 17.38 & 17.03 & 15.41 & \textbf{30.33} & 17.82 \\
Ophthalmology and Sensory Systems & 33.29 & 31.45 & 25.10 & \underline{38.98} & 22.46 & 24.66 & 23.78 & 22.20 & \textbf{41.86} & 25.18 \\
Orthopedics and Musculoskeletal & 6.27 & 5.99 & 5.25 & \underline{6.86} & 4.77 & 5.62 & 4.57 & 5.57 & \textbf{12.31} & 5.38 \\
Pharmacology & 37.47 & 35.19 & 29.42 & \underline{41.86} & 25.59 & 27.52 & 26.99 & 25.01 & \textbf{44.43} & 30.80 \\
Psychiatry and Mental Health & 39.71 & 39.69 & 28.71 & \underline{40.74} & 34.21 & 28.06 & 33.99 & 29.43 & \textbf{50.28} & 35.40 \\
Pubmed & \underline{68.72} & 56.90 & 58.72 & 67.32 & 29.90 & 37.75 & 29.47 & 22.19 & \textbf{70.06} & 41.53 \\
Radiology and Imaging & 47.39 & 40.43 & 41.84 & \underline{51.11} & 30.95 & 36.64 & 19.80 & 26.58 & \textbf{52.86} & 32.25 \\
Reproductive System & 32.05 & 32.25 & 25.87 & \underline{39.40} & 25.09 & 28.28 & 29.29 & 24.48 & \textbf{42.09} & 28.66 \\
Respiratory and Pulmonology & 44.15 & 39.25 & 34.76 & \underline{49.31} & 28.89 & 31.80 & 31.56 & 26.25 & \textbf{50.43} & 2.74 \\
Surgical Specialties & 38.07 & 36.10 & 34.06 & \underline{46.42} & 25.42 & 32.23 & 27.90 & 26.93 & \textbf{48.08} & 31.71 \\
\midrule
\textbf{Average} & 38.07 & 35.00 & 32.00 & \underline{43.00} & 26.00 & 28.00 & 26.00 & 25.00 & \textbf{45.00} & 24.80 \\
\bottomrule
\end{tabular}

\caption{NDCG@10 scores for ten retrieval models representing different retrieval styles, including \ccr{lexicon-based retrievers}, \ccb{uni-modal dense retrievers}, \ccg{CLIP-style models}, \ccy{multi-modal dense retrievers}, and \ccm{late-interaction multi-modal retrievers} for the Visual Context Retrieval task in the {\it M3Retrieve} Benchmark. The best-performing model has been highlighted as \textbf{bold} while the second best model has been \underline{underlined}.}
\label{tab:results_visualcontext}
\end{table*}
Based on the results in Table \ref{tab:results_visualcontext} we can make the following conclusions : \par
 \textbf{1) Anatomy and Physiology:} NV Embed (66.29) and BM25 (61.20) outperform all other models, showing that \textbf{dense text retrieval remains the most effective method for this domain}. \par
     \textbf{2) Psychiatry and Mental Health:} \textbf{MM Embed (50.28)} is the strongest performer, demonstrating the benefit of \textbf{multimodal representation learning} in capturing mental health-related contexts.\par
     \textbf{3) PubMed Retrieval:} \textbf{MM Embed (70.06)} surpasses \textbf{BM25 (68.72)}, highlighting that multimodal encoders provide a \textbf{more comprehensive representation} in large-scale biomedical literature retrieval.\par
     \textbf{4) Orthopedics and Musculoskeletal:} Scores remain \textbf{low across all models} (BM25: 6.27, NV Embed: 6.86), indicating \textbf{retrieval challenges} in this sub-domain, possibly due to \textbf{complex terminologies and limited training data}.\par

Overall, \textbf{dense encoders} (MM Embed, NV Embed) outperform both \textbf{BM25} and \textbf{CLIP-style models}, highlighting the benefits of \textbf{deep learning-based joint embeddings}. \textbf{BM25 remains a strong baseline}, particularly in text-heavy disciplines.  

The superior performance of MM Embed over all unimodal models, including NV Embed, the best-performing text retrieval model, emphasizes the \textbf{ importance of multimodal representation learning} for medical retrieval in \textit{M3Retrieve}. However, NV Embed and MM Embed show \textbf{30\% and 25.37\% lower} NDCG@10 scores than their BEIR averages (62.65 and 60.3).

\subsubsection{Case Study Retrieval}

\begin{table*}[!htb]
    \centering
    \tiny

\begin{tabular}{l|%
\ccrcol|%
\ccbcol\ccbcol\ccbcol|%
\ccgcol\ccgcol\ccgcol|%
\ccycol\ccycol|%
\ccmcol}  
\toprule
Discipline & BM-25 & E5 & BGE & NV Embed & MM Ret & MII & CLIP SF & BLIP FF & MM Embed & FLMR \\
\midrule
Obstetrics and Gynecology        & 13.90 & 11.14 &  9.34 &  8.12 & 1.69 & 1.65 & 1.10 & 0.80 & 14.47 & 1.55 \\
Hematology                       & 10.32 &  6.33 &  5.62 &  9.02 & 1.16 & 1.06 & 1.04 & 1.05 &  7.84 & 0.80 \\
Cardiology                       & 10.09 &  6.76 &  6.92 &  9.88 & 0.98 & 2.31 & 2.08 & 0.58 &  8.98 & 2.30 \\
Neurology                        & 11.16 &  7.82 &  6.11 &  9.55 & 0.98 & 2.02 & 1.35 & 0.90 &  8.84 & 1.35 \\
Orthopedics                      & 14.21 & 10.39 &  9.09 & 11.23 & 1.29 & 2.70 & 1.50 & 1.10 &  9.90 & 0.95 \\
Pathology                        & 13.51 &  6.96 &  6.17 & 12.67 & 0.87 & 1.87 & 1.20 & 0.85 & 10.05 & 2.10 \\
Endocrinology                    & 11.14 &  7.53 &  5.73 &  9.97 & 0.95 & 1.39 & 0.81 & 1.01 & 10.01 & 1.40 \\
Oncology                         & 10.16 &  6.10 &  5.33 & 13.04 & 0.75 & 1.43 & 0.90 & 0.70 &  7.34 & 0.60 \\
Pulmonology                      & 11.07 &  6.88 &  6.54 &  8.98 & 0.90 & 1.43 & 1.05 & 1.00 &  9.03 & 1.45 \\
Psychiatry and Behavioral Health &  7.18 &  5.58 &  4.18 & 11.98 & 0.51 & 0.86 & 1.40 & 0.95 &  8.48 & 1.10 \\
Genetics and Genomics            & 12.33 &  8.37 &  7.13 & 13.59 & 1.58 & 1.66 & 1.13 & 0.88 & 12.13 & 2.45 \\
Infectious Diseases              & 12.08 &  8.06 &  6.45 & 11.48 & 1.22 & 1.31 & 1.18 & 0.92 &  9.90 & 0.75 \\
Gastroenterology                 & 13.08 &  8.84 &  7.99 & 12.11 & 1.30 & 1.95 & 1.35 & 1.10 & 11.41 & 2.05 \\
Rheumatology and Immunology      & 10.18 &  6.71 &  5.15 & 12.30 & 0.87 & 0.84 & 1.85 & 0.79 &  9.45 & 1.30 \\
Dermatology                      & 12.16 &  7.74 &  6.84 & 10.93 & 1.25 & 2.15 & 1.26 & 1.17 & 10.75 & 2.05 \\
\midrule
\textbf{Average}                & 11.50 &  7.68 &  6.59 & 10.99 & 1.09 & 1.64 & 1.28 & 0.92 &  9.91 & 1.48 \\
\bottomrule
\end{tabular}
\caption{NDCG@10 scores for ten retrieval models representing different retrieval styles, including \ccr{lexicon-based retrievers}, \ccb{uni-modal dense retrievers}, \ccg{CLIP-style models}, \ccy{multi-modal dense retrievers}, and \ccm{late-interaction multi-modal retrievers} for the Case study Retrieval task in the {\it M3Retrieve} Benchmark.}
\label{tab:results_case}
\end{table*}

According to Table \ref{tab:results_case}, across the fifteen medical specialties, traditional term‐matching via BM25 remains a strong baseline, especially in Orthopedics (14.21), Pathology (13.51), and Gastroenterology (13.08). Neural text encoders (E5 and BGE) demonstrate particular strength in Gastroenterology (E5 = 8.84, BGE = 7.99) and Genetics and Genomics (E5 = 8.37, BGE = 7.13), suggesting their aptitude for capturing nuanced biomedical language. The NV Embed model shows a clear niche in Genetics and Genomics (13.59), while multimodal retrieval (MM Ret) and the Medical Image Integrator (MII) offer modest yet consistent improvements (averages of 1.09 and 1.64, respectively), with MII peaking in Cardiology (2.31). Vision–language adapters, CLIP SF and BLIP FF, deliver complementary gains: CLIP SF excels in Rheumatology and Immunology (1.85) and Cardiology (2.08), and BLIP FF adds notable lift in Dermatology (1.17). Finally, the unified FLMR model achieves robust performance across domains, reaching its highest scores in Genetics and Genomics (2.45) and Gastroenterology (2.05), underscoring its versatility for image‐informed medical retrieval.

\subsubsection{Query to Image Retrieval}
\begin{table*}[!htb]
    \centering
    \tiny

\begin{tabular}{l|%
\ccgcol\ccgcol\ccgcol|%
\ccycol\ccycol|%
\ccmcol}  
\toprule
Discipline & MM Ret & MII   & CLIP SF & BLIP FF & MM Embed & FLMR \\
\midrule
Spine and Muscles                  & 3.50 & 37.88 & 28.65 & 6.00 & 26.01 & 2.10 \\
Abdomen                            & 1.45 & 49.04 & 29.41 & 1.22 & 27.54 & 3.00 \\
Head                               & 2.05 & 34.31 & 19.38 & 1.69 & 26.54 & 2.50 \\
Thorax                             & 3.38 & 47.27 & 33.75 & 1.00 & 35.03 & 2.80 \\
Reproductive and Urinary System    & 1.01 & 49.14 & 34.13 & 1.28 & 32.36 & 2.40 \\
\midrule
\textbf{Average}                   & 2.28 & 43.53 & 29.12 & 2.07 & 29.49 & 2.56 \\
\bottomrule
\end{tabular}

\caption{NDCG@10 scores for ten retrieval models representing different retrieval styles, including \ccr{lexicon-based retrievers}, \ccb{uni-modal dense retrievers}, \ccg{CLIP-style models}, \ccy{multi-modal dense retrievers}, and \ccm{late-interaction multi-modal retrievers} for the Query to Image Retrieval task in the {\it M3Retrieve} Benchmark.}
\label{tab:IT2I}
\end{table*}

\section*{Domain-wise Model Performance Analysis}

Table~\ref{tab:IT2I} presents the performance of various models across different anatomical domains, namely Spine and Muscles, Abdomen, Head, Thorax, and the Reproductive and Urinary System. 

In the \textbf{Spine and Muscles} domain, the highest performance is observed with the MII model (37.88), while CLIP SF and MM Embed also demonstrate competitive performance (28.65 and 26.01, respectively). BLIP FF shows moderate alignment (6.00), while MM Ret and FLMR report lower scores (3.50 and 2.10).

In the \textbf{Abdomen}, MII achieves the strongest result (49.04), followed by CLIP SF (29.41) and MM Embed (27.54). Performance from BLIP FF (1.22), MM Ret (1.45), and FLMR (3.00) remains modest.

The \textbf{Head} domain reveals comparatively lower performance across most models. MII again leads (34.31), while MM Embed and CLIP SF register similar outcomes (26.54 and 19.38, respectively). Other models exhibit limited effectiveness.

For the \textbf{Thorax}, both MII (47.27) and MM Embed (35.03) perform strongly, with CLIP SF also maintaining a good score (33.75). However, MM Ret (3.38), BLIP FF (1.00), and FLMR (2.80) show relatively lower results.

Lastly, in the \textbf{Reproductive and Urinary System}, MII (49.14) and CLIP SF (34.13) dominate, while MM Embed also performs reasonably well (32.36). The remaining models perform below par in this category.

\textbf{Overall}, MII consistently achieves the highest average performance (43.53), indicating robust multi-domain capabilities. CLIP SF (29.12) and MM Embed (29.49) also show strong generalization. In contrast, BLIP FF (2.07), MM Ret (2.28), and FLMR (2.56) underperform on average, indicating more limited domain adaptation.

\end{document}